\documentclass[12pt,twoside]{article}
\usepackage{amsmath,amsfonts,amssymb}
\usepackage{latexsym}
\usepackage{dcolumn}
\usepackage{graphicx,epsfig}
\usepackage{amsthm}
\usepackage{hyperref}

\begin{document}
\begin{center}
{\Large\bf Perfect Fluid Quantum Anisotropic Universe: Merits and Challenges}
\\[15mm]
Barun Majumder{\footnote{E-mail: barunbasanta@iiserkol.ac.in}} and Narayan Banerjee{\footnote {E-mail: narayan@iiserkol.ac.in}}\\ 
Department of Physical Sciences, \\
Indian Institute of Science Education and Research - Kolkata,\\
Mohanpur Campus, P.O. BCKV Main Office, District Nadia,\\ 
West Bengal 741252, India.
\end{center}
\date{}
\vspace{0.5cm}
\begin{abstract}
The present paper deals with quantization of perfect fluid anisotropic cosmological models. Bianchi type V and IX models are discussed following Schutz's method of expressing fluid velocities in terms of six potentials. The wave functions are found for several examples of equations of state. In one case a complete wave packet could be formed analytically. The initial singularity of a zero proper volume can be avoided in this case, but it is plagued by the usual problem of non-unitarity of anisotropic quantum cosmological models. It is seen that a particular operator ordering alleviates this problem.
\end{abstract}
{\em PACS numbers: 04.20.Cv., 04.20.Me}
\section{Introduction}
The basic motivation behind a  quantum cosmological model is two-fold. One is the fact that when the linear dimension of the universe reaches the Planck scale ($~10^{-33}$cm), the universe is indeed governed by a quantum picture. The second is the hope that a quantum description might be able to produce a singularity free birth of our universe. The quantum state of the universe is described by a wave function  $\Psi$ which is the solution of the Wheeler DeWitt equation on a minisuperspace. For a comprehensive review, we refer to \cite{swh}. \\
\par One critical problem in quantum cosmology is certainly that of a suitable choice of time against which the evolution of the universe is investigated. This is because the notion of time has different implications in General Relativity and Quantum Mechanics. Systematic attempts towards resolving this problem started in the 90's\cite{ku}. The strategies also include a scenario where the notion of time plays no role whatsoever\cite{carlo}. For a recent review of different strategies, see \cite{ander}. \\
\par If matter is taken as a perfect fluid, the strategy adopted by Schutz\cite{r3,r4} becomes extremely useful as a set of canonical transformations leads to one conjugate momentum associated with the fluid giving a linear contribution to the Hamiltonian. The corresponding fluid variable thus qualifies to play the role of time in the relevant Schr$\ddot{o}$dinger equation. In an ever expanding model, the fluid density has a monotonic temporal behaviour, the time orientability is thus ensured. \\
\par Schutz's formalism has been extensively used by Alvarenga {\it et al}\cite{alvarenga1} for isotropic cosmological models. The investigation involves spatially flat, open and closed models. Matter content  is taken as a perfect fluid with an equation of state $p=\alpha \rho$. They found finite norm wave packet solutions of the Wheeler-DeWitt equations. One important finding is that some singularity free models could be constructed even without violating the energy conditions. A later work involves anisotropic Bianchi I cosmology\cite{r11}, which is the anisotropic generalization of a spatially flat isotropic model. Anisotropic models have problems, particularly that of non-unitarity. The Hamiltonian, although hermitian, is not self-adjoint. This is  discussed in some detail in ref\cite{r11}.\\
\par It is well known that a standard Copenhagen interpretation of a quantum cosmological model is not tenable. A ``Bohm-de Broglie'' interpretation  normally  does better. For a comprehensive review, we refer to \cite{neto}. In the anisotropic case, the non unitarity leads to futher problems in the interpretation. As the norm of the wave function becomes time dependent, Bohmian trajectories are not conserved and the Bohm-de Broglie interpretation also becomes vulnerable\cite{r11}. It appears that the philosophy of interpretation might require a dramatic extension so as to incorporate anisotropic cosmologies. 
\par In the present work Schutz's formalism is utilized to quantize Bianchi type V and type IX perfect fluid cosmological models which are the anisotropic generalzations of open and closed isotropic models respectively. In section 2 we discuss the formalism and use that to quantize  Bianchi type V models. In section 3, a type IX model is quantized. Some concluding remarks are made in section 4.

\section{Schutz's formalism and quantization of Bianchi V cosmological model}
The relevant action for gravity with a perfect fluid can be written as
\begin{equation}
{\cal A} = \int_Md^4x\sqrt{-g}\,R + 2\int_{\partial M}d^3x\sqrt{h}\, h_{ab}\, K^{ab} + \int_Md^4x\sqrt{-g}\,\, P
\end{equation}
where $h_{ab}$ is the induced metric over three dimensional spatial hypersurface which is the boundary $\partial M$ of the four dimensional manifold $M$ and $K^{ab}$ is the extrinsic curvature. Here units are so chosen that $c=16\pi G=\hbar$ is equal to one. $P$ is the pressure due to the perfect fluid. The perfect fluid satisfies an equation of state
\begin{equation}
P=\alpha \rho,
\end{equation}
where $\alpha\leq1$. This restriction stems from the consideration that sound waves cannot propagate faster than light. In Schutz's formalism \cite{r3,r4} the fluid's four velocity can be expressed in terms of six potentials. However, two of them are connected with rotation. In Bianchi V or IX models permit timelike geodesics which are hypersurface orthogonal, the rotation tensor $\omega_{\mu \nu}$ vanishes, and one can write the four velocity in terms of only four independent potentials as
\begin{equation}
u_\nu = \frac{1}{h}(\epsilon_{,\nu} + \theta S_{,\nu}).
\end{equation} 
Here $h$, $S$, $\epsilon$ and $\theta$ are the velocity potentials having their own evolution equations, where the potentials connected with vorticity are dropped. The four velocity is normalized as 
\begin{equation}\label{nor}
u^\nu u_\nu = 1.
\end{equation}
Although the physical identification of velocity potentials are irrelevant for the formulation, $h$ and $S$ can be identified with the specific enthalpy and specific entropy respectively. This identification facilitates the representation of fluid parameters in terms of thermodynamic quantities.
\par
The metric for the Bianchi V anisotropic model is written as
\begin{equation}
ds^2 = n^2(t)dt^2 - a^2(t)dx^2 - e^{2mx}\,[\,b^2(t)dy^2 + c^2(t)dz^2\,] 
\end{equation}
where $n(t)$ is called the lapse function. While $a, b, c$ are functions of the cosmic time $t$, $m$ is a constant. Bianchi type I model is recovered when $m=0$. Eliminating the surface terms, the first and second terms of the action (1) give
\begin{equation}
{\cal A}_g = \int dt\bigg[-\frac{2}{n}\bigg(\dot a\dot bc + \dot a\dot cb +
\dot b\dot ca + 3n^2m^2 \frac{bc}{a} \bigg)\bigg] \, ,
\end{equation}
where an overhead dot indicates a differentiation with respect to time $t$ and ${\cal A}_g$ is contribution of geometry to the action.
So the gravitational Lagrangian density can be easily identified as
\begin{equation}
{\cal L}_g = -\frac{2}{n}\big(\dot a\dot bc + \dot a\dot cb + \dot b\dot ca\big) - 6nm^2 \frac{bc}{a}  \, .
\end{equation}
If we now choose the three metric coefficients as
\begin{equation}
a(t) = e^{\beta_0} \, ,\, b(t) = e^{\beta_0 + \beta_+ - \beta_-} \, ,\, c(t) = e^{\beta_0 - \beta_+ + \beta_-} \,
\end{equation}
then we get
\begin{equation}
a(t)\,b(t)\,c(t) = e^{3\beta_0} \, 
\end{equation}
and
\begin{equation}
{\cal L}_g = -2\frac{e^{3\beta_0}}{n}\big[3\,\dot {\beta_0}^2 - \big(\dot {\beta_+} - \dot {\beta_-}\big)^2\,\big] - 6e^{\beta_0}nm^2 \,  .
\end{equation}
Now $\beta_0$, $\beta_+$ and $\beta_-$ will be treated as the relevant variables in place of $a$, $b$ and $c$. The corresponding conjugate momenta are then
\begin{align}
\label{mom}
& p_0 = - 12\frac{e^{3\beta_0}}{n}\dot{\beta_0} \, , \nonumber \\ & p_+ = 4\frac{e^{3\beta_0}}{n}(\dot{\beta_+} - \dot{\beta_-}) \,, \nonumber \\ 
& p_- = - 4\frac{e^{3\beta_0}}{n}(\dot{\beta_+} - \dot{\beta_-})  \,.
\end{align}
The Hamiltonian of the gravity sector is now given by
\begin{equation}
{\cal H}_g = ne^{-3\beta_0} \bigg(-\frac{p_0^2}{24} + \frac{p_+^2}{8} + 6e^{4\beta_0}m^2 \,\bigg),
\end{equation}
where momentum $ p_-$ is replaced in terms of  $p_+$ using equation (\ref{mom}).
For the fluid part, the action can be written using thermodynamic relations for $h$ and $S$ \cite{r3,r4}.
The relevant equations are 
\begin{equation}
\label{thermo}
\rho = \rho_0(1+\Pi)\quad , \quad h = 1+\Pi+P/\rho_0 \quad , \quad \tau dS = d\Pi+P d(1/\rho_0) \quad
\end{equation}
where $\tau$, $\rho$, $\rho_0$ and $\Pi$ are temperature, total mass energy density, rest mass density and specific internal energy respectively. Rewriting the third equation of (\ref{thermo}) we get
\begin{equation}
\tau dS =(1+\Pi)\,d\left[\mbox{ln}(1+\Pi)-\alpha \mbox{ln} \rho_0\right] \quad .
\end{equation}
It then follows that, $\tau=1+\Pi$ and $S=\mbox{ln}(1+\Pi)-\alpha \mbox{ln} \rho_0$. We can show that the equation of state takes the form
\begin{equation}
P=\frac{\alpha}{{(1+\alpha)}^{1+1/\alpha}}h^{1+1/\alpha}e^{-S/\alpha} \, .
\end{equation}
In a comoving system $u_{\nu}=(n,0,0,0)$, and equation (\ref{nor}) yields 
\begin{equation}\label{af}
{\cal A}_f = \int dt\biggr[n^{-1/\alpha}e^{3\beta_0}\frac{\alpha}{(1+\alpha)^{1+1/\alpha}}(\dot\epsilon + \theta\dot S)^{1+1/\alpha}e^{-S/\alpha} \biggl] \, ,
\end{equation}
${\cal A}_f$ being the fluid part of the action.
As $h>0$ so $(\dot\epsilon + \theta\dot S)>0$. If we try the canonical methods used, for example, in \cite{r5} the Hamiltonian for this action can be written in a very simple form with the canonical transformations
\begin{equation}
T = p_Se^{-S}p_\epsilon^{-(1 + \alpha)} \quad , \quad p_T = p_\epsilon^{1+\alpha}e^S \quad , \quad
\bar\epsilon = \epsilon - (1 + \alpha)\frac{p_S}{p_\epsilon} \quad , \quad \bar p_\epsilon = p_\epsilon \quad ,
\end{equation}
along with $p_S=\theta p_\epsilon$. Here $p_\epsilon =\frac{\partial {\cal L}_f}{\partial\dot\epsilon}$, $p_S =\frac{\partial {\cal L}_f}{\partial\dot S}$ and ${\cal L}_f$, the Lagrangian density of the fluid, is the expression inside the square bracket of equation (\ref{af}). The Hamiltonian for this perfect fluid can now be written as
\begin{equation}
{\cal H}_f=ne^{-3\beta_0}e^{3(1-\alpha)\beta_0}p_T \, .
\end{equation}
The advantage of using this method, i.e., using canonical transformations, is that we could find a set of variables where the system of equations is more tractable, while the Hamiltonian structure of the system remains intact. It also deserves mention that amongst the four velocty potentials mentioned, actually two are used, namely $\epsilon$ and $S$. This is because $\epsilon$ and $h$ are related by $$u^\mu {\epsilon}_{,\mu} = - h $$ ( see ref\cite{r3}) and one other, namely $\theta$ can be settled using the normalization (\ref{nor}).

The super Hamiltonian for the minisuperspace of this anisotropic quantum model is
\begin{align}
\label{supham}
{\cal H} &= {\cal H}_g + {\cal H}_f \nonumber \\ 
&= n\frac{e^{-3\beta_0}}{24}\big[-p_0^2 + 3\, p_+^2 + 144\, e^{4\beta_0}m^2 + 24\, e^{3(1-\alpha)\beta_0}p_T \,\big] \, .
\end{align}
Here $n$ acts as a Lagrange multiplier taking care of the classical constraint equation ${\cal H}=0$. Using the usual quantization procedure \cite{r6,r7}, we write the Schr$\ddot{o}$dinger-Wheeler-DeWitt equation for our super Hamiltonian with the ansatz that the super Hamiltonian operator annihilates the wave function,
\begin{equation}
\label{wdwe1}
\hat{{\cal H}}\quad \vert\Psi(\beta_0,\beta_+,t)\,\rangle=0.
\end{equation}
 There are attempts\cite{rosen} to show that the classical Hamiltonian for a cosmological spacetime is zero, if one takes into account both the matter sector and the geometry sector together, like the present situation. But these attempts involves pseudotensorial calculations and the result depends on the minisuperspace chosen\cite{nbss}. The problem perhaps stems from the fact that localization of energy in general relativity is not uniquely defined. Whatever be the status of the constraint (\ref{wdwe1}) in the most general case, we shall be using this following the standard practice. \\
With $p_{x_i}\rightarrow -i\partial_{x_i}$, $p_T \rightarrow i\partial_T$, $\hbar=1$ equation (\ref{wdwe1}) can now be written as 
\begin{equation}
\label{wdwe}
\bigg[\frac{\partial^2}{\partial \beta_0^2} - 3\,\frac{\partial^2}{\partial \beta_+^2} + 144\,e^{4\beta_0}\,m^2 +
24\,i\,e^{3(1-\alpha)\beta_0}\,\frac{\partial}{\partial T}\bigg]\, \Psi(\beta_0,\beta_+,t) = 0 \, \, .
\end{equation}
In this equation $T=t$ is the cosmic time co-ordinate if we choose the gauge $n=e^{3\alpha \beta_0}$ and this follows from Hamilton's classical equations as $\dot T=\{T,{\cal H}\}=\frac{n}{e^{3\alpha \beta_0}}$. Now it must be mentioned that while constructing the Schr$\ddot{o}$dinger-Wheeler-DeWitt equation (\ref{wdwe}) we have considered a particular choice of factor ordering for $p_0$ and $e^{-3\beta_0}$. We will discuss other choices of factor orderings and its consequences on our prime results in the subsequent section. We require that the super Hamiltonian must be hermitian, so the wave function $\Psi$ must satisfy these
conditions \cite{r12,r13,r11};
\begin{equation}
\label{bc}
\Psi\arrowvert_{x_i \rightarrow \pm \infty}=0 \,\,\,\, \text{or} \,\,\,\, \Psi'\arrowvert_{x_i \rightarrow \pm \infty}=0 \, .
\end{equation}
In order to solve for the wave function $\Psi$ from equation (\ref{wdwe}) we employ a  separation of variables as, 
\begin{equation}
\Psi (\beta_0, \beta_+, T) = e^{-iET}\, \xi (\beta_0,\beta_+) \, ,
\end{equation}
we get
\begin{equation}
\frac{\partial^2 \xi}{\partial \beta_0^2} - 3 \frac{\partial^2 \xi}{\partial \beta_+^2} + 144\,e^{4\beta_0}\,m^2 \xi + 24 \, E \, 
e^{3(1-\alpha)\beta_0} \xi = 0 \,,
\end{equation}
where $E$ is the separation constant. Further if we write
\begin{equation}
\xi (\beta_0, \beta_+) = \phi (\beta_0) \eta (\beta_+)\,,
\end{equation}
we get
\begin{equation}
\frac{1}{\phi}\frac{\partial^2 \phi}{\partial \beta_0^2} - \frac{3}{\eta} \frac{\partial^2 \eta}{\partial \beta_+^2} + 
144\,e^{4\beta_0}\,m^2 + 24 \, E \, e^{3(1-\alpha)\beta_0} = 0 \,.
\end{equation}
The solution for $\eta$ is
\begin{equation}
\eta (\beta_+) = C\,e^{i\tfrac{k}{\sqrt{3}}\beta_+} ~,
\end{equation}
where $C$ is the integration constant and $k$ is again a constant of separation which has to be real so that the wave function is normalizable. The equation
for $\phi$ is
\begin{equation}
\label{vs1}
\frac{\partial^2 \phi}{\partial \beta_0^2} + 144\,e^{4\beta_0}\,m^2 \phi + 24 \, E \, e^{3(1-\alpha)\beta_0} \phi + k^2 \phi = 0 ~.
\end{equation}
For $m=0$, this equation reduces to the corresponding equation for the Bianchi I model \cite{r11}. It is very difficult to solve this equation analytically for all allowed values of $\alpha$. Here we will study the solution for some particular values of $\alpha$.
\subsection{${\bf \alpha = -\frac{1}{3}}$ (Distribution of Strings)}
With $\alpha=-1/3$ equation (\ref{vs1}) becomes 
\begin{equation}
\label{phi1}
\frac{\partial^2 \phi}{\partial \beta_0^2} + 144\,e^{4\beta_0}\,m^2 \phi + 24 \, E \, e^{4\beta_0} \phi + k^2 \phi = 0 ~.
\end{equation}
The solution of equation (\ref{phi1}) is known in terms of Bessel function and now we can write the wave function of equation (\ref{wdwe}) as \cite{bell}
\begin{equation}
\label{comp1}
\Psi (\beta_0, \beta_+, T) = e^{-iET}~e^{i\tfrac{k}{\sqrt{3}}\beta_+} \big[ C_1 J_{\frac{ik}{2}} \big(\sqrt{36m^2+6E}~e^{2\beta_0}\big)
+ C_2 J_{\frac{-ik}{2}} \big(\sqrt{36m^2+6E}~e^{2\beta_0}~\big)~\big] ~,
\end{equation}
where $C_1$, $C_2$ are the integration constants. Now we can construct a regular wave packet superposing these solutions. In principle this can be easily done
by considering the arbitrary integration constants to be suitable Gaussian functions of the parameters $k$ and $E$ ( see ref \cite{r11}). Defining $q = \sqrt{36m^2+6E}$ we
can write the form of the wave packet as
\begin{equation}
\Psi_{wp} = \int_{k=-\infty}^\infty ~ \int_{q=0}^\infty ~ f(k,q)~q~ e^{i6m^2T}~ e^{-i\frac{q^2}{6}T}~ e^{i\frac{k}{\sqrt{3}}\beta_+}~
J_{\frac{ik}{2}}(qe^{2\beta_0})~ dk~dq~~,  
\end{equation} 
where $f(k,q) = 2^{\frac{ik}{2}+1}~q^{\frac{ik}{2}}~ e^{-\lambda q^2 - \gamma k^2}$ and $\lambda$ and $\gamma$ are arbitrary positive constants. The
above integrals can be explicitly evaluated and the wave packet becomes
\begin{equation}
\Psi_{wp} = \sqrt{\pi/\gamma}~ e^\theta ~ e^{i6m^2T} ~ e^{-\frac{e^{4\beta_0}}{4(\lambda + \frac{iT}{6})}} ~ 
e^{-\frac{(\beta_0 + \frac{\beta_+}{\sqrt{3}} + \frac{\theta}{2})^2}{4\gamma}} ~~,
\end{equation}
where $\theta = - \ln ~ (\lambda + \frac{iT}{6})$. The norm of the wave packet is 
\begin{equation}
\label{wnorm}
\int_{-\infty}^\infty ~ \int_{\infty}^\infty ~ e^{4\beta_0}~ \Psi_{wp}^* ~\Psi_{wp} ~ d\beta_0 ~ d\beta_+ = 
\tfrac{\pi}{2\lambda}~ \sqrt{\tfrac{6\pi}{\gamma}} ~ e^{\frac{\omega^2}{8\gamma}} ~~,
\end{equation}
where $\omega = \arctan \frac{T}{6\lambda}$. Clearly we can see that the norm is time dependent and hence the model is not unitary. This is not surprising as
we know that the anisotropic quantum cosmological models are non-unitary \cite{r11}. The expectation value of any variable $\beta_i$ can now be calculated as
\begin{equation}
\langle \beta_i \rangle = \frac{\int_{-\infty}^\infty ~ \int_{\infty}^\infty ~ e^{4\beta_0}~ \Psi_{wp}^*~\beta_i ~\Psi_{wp} ~ d\beta_0 ~ d\beta_+}{\int_{-\infty}^\infty ~ \int_{\infty}^\infty ~ e^{4\beta_0}~ \Psi_{wp}^* ~\Psi_{wp} ~ d\beta_0 ~ d\beta_+} ~~.
\end{equation}
For $\beta_i = \beta_0$ we find
\begin{equation}
\label{e35}
\langle \beta_0 \rangle = - \tfrac{1}{4}~ (\ln \tfrac{\lambda}{2B^2} + l) ~~,
\end{equation}
where $B = \sqrt{\lambda^2 + \tfrac{T^2}{36}}$ and $l$ is a numerical factor $\sim 0.58$. Now
\begin{equation}
e^{\langle \beta_0 \rangle} = (a~b~c)^{1/3} = (2\lambda e^{-l})^{\frac{1}{4}} ~ (1 + \tfrac{T^2}{36\lambda^2})^{1/4} ~~.
\end{equation}
This describes the cosmological evolution of the spatial volume and this model predicts a bounce from a minimum volume universe with no singularity. Of course this result is facilitated by the choice of the gauge $n= e^{3\alpha{\beta}_0}$ as mentioned after equation(21). It is easy to see that the minimum volume is obtained at $T=0$ and the corresponding length scale is $(2\lambda e^{-l})^{\frac{1}{4}}$. as already mentioned, $l=0.58$. So the linear size of the universe can be set in the model by choosing the vaule of $\lambda$. For $\beta_i = \beta_+$,
\begin{equation}
\label{e37}
\langle \beta_+ \rangle = -\tfrac{1}{4}~ (\ln \tfrac{\lambda}{2} + l).
\end{equation}
It deserves mention that $\beta_+$ is in fact a constant in time. 
\begin{figure}[!h]
\hspace{-.5cm} \centerline{\psfig{figure = 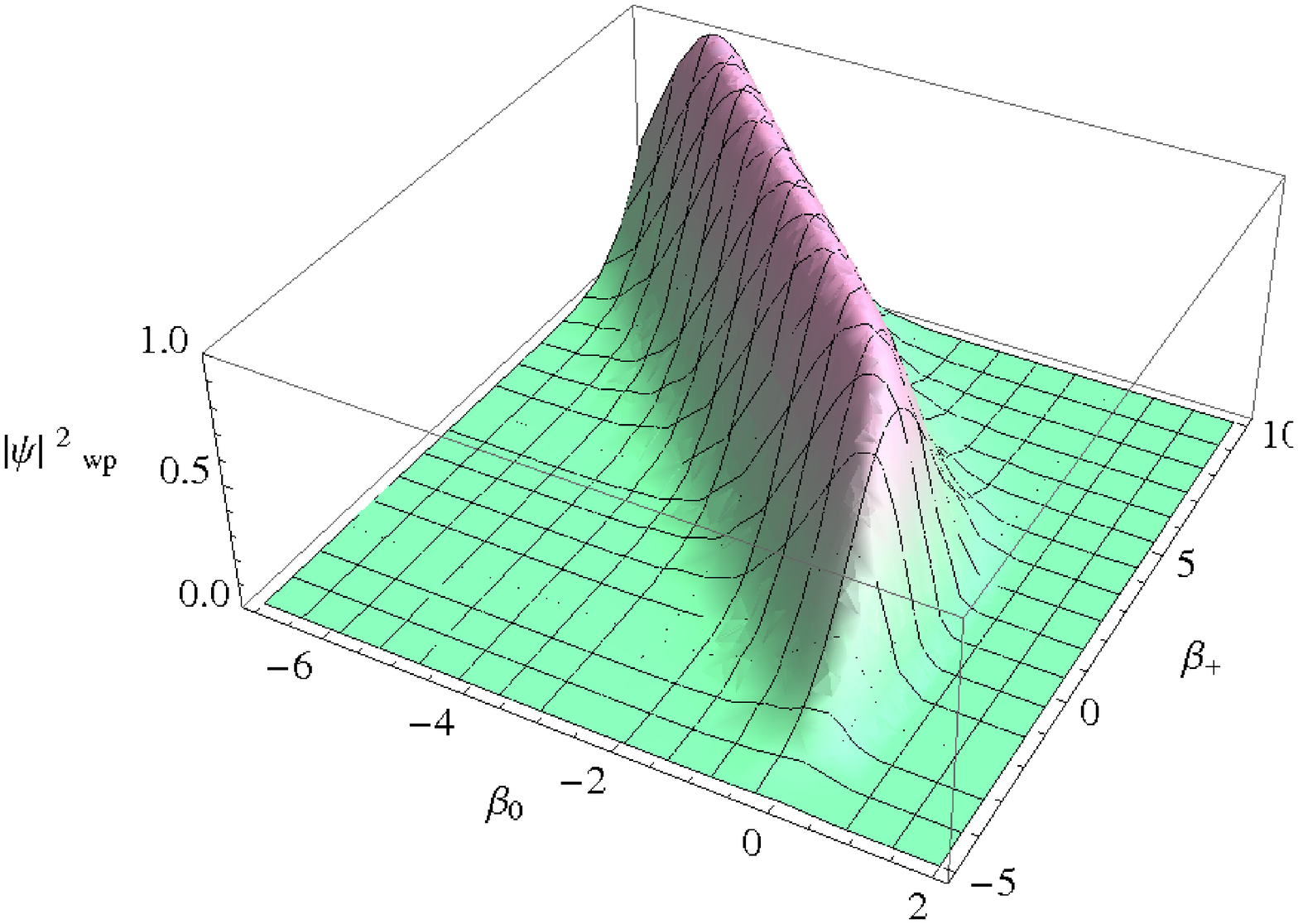, height=70mm, width=100mm} \hskip 5pt \psfig{figure = 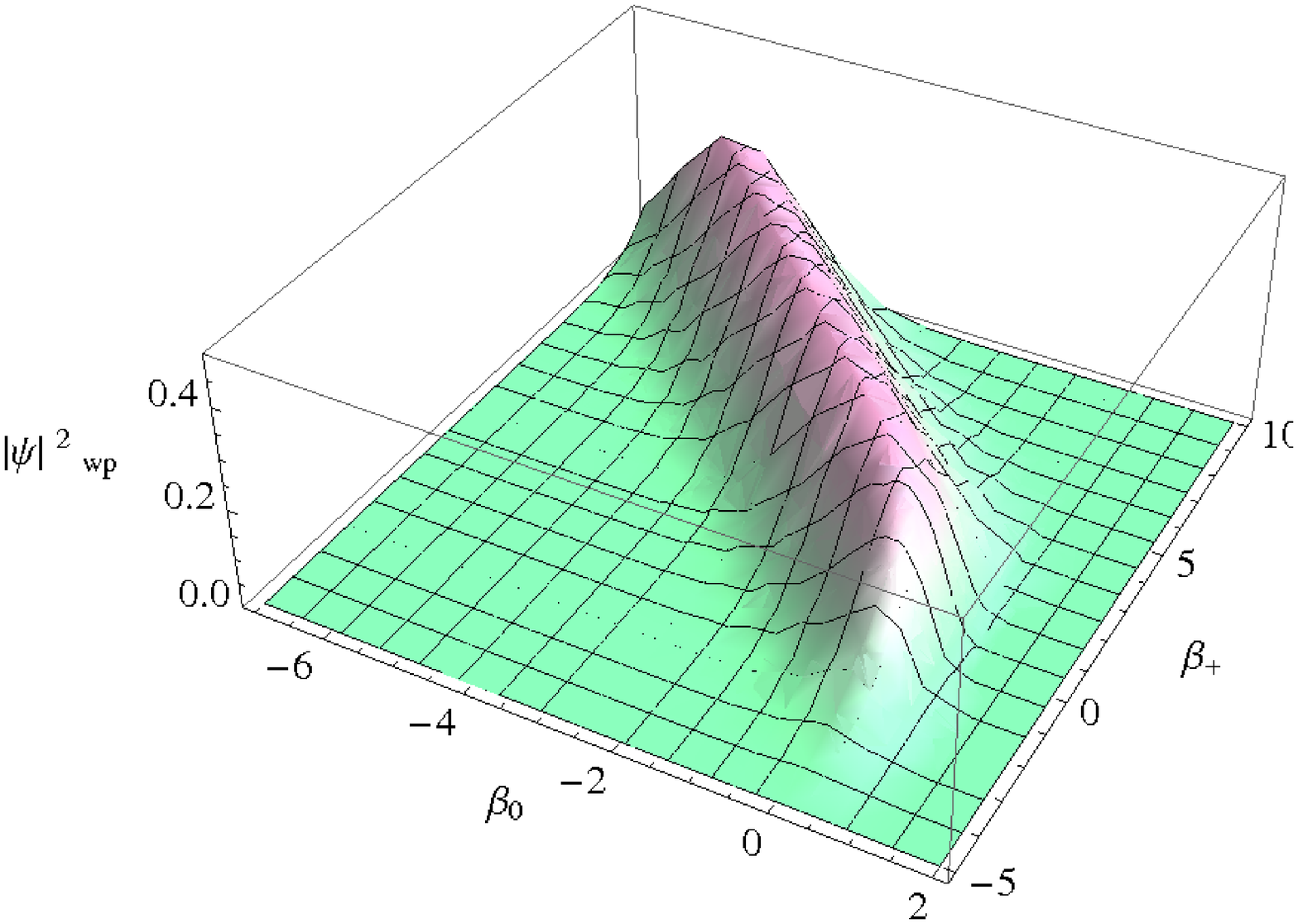, height=70mm, width=100mm}}
\caption{\scriptsize{Plot of $|\Psi|^{2}$ against $\beta_0$ and $\beta_+$ at a fixed time. The left panel shows the plot at $T=0$ and the right panel is the plot for $T=10$.}}
\label{fig3}
\end{figure}
The plot of the norm of the wave packet against $\beta_0$ and $\beta_+$ shows a well behaved pattern, and even at $T=0$, does not show any sign of blowing up (see Figure [\ref{fig3}]). This is consistent with the fact that there exists a minimum of the proper volume and thus the pathology of a singular state of the universe can be avoided in this model. Furthermore, the plots are given for two different time, namely $T=0$ and $T=10$ in some units, and the qualitative nature of the plots remain similar.
\par
\begin{figure}[htb]
\begin{tabular}{c}
\hspace{2.3cm} \includegraphics[width=11cm,height=8cm]{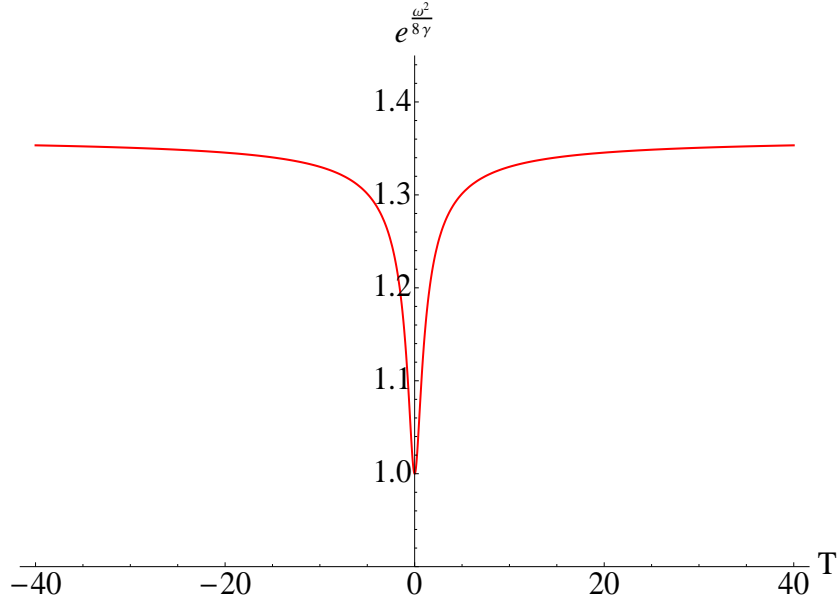}
\end{tabular}
\caption{\scriptsize Plot of Norm ($\tfrac{\pi}{2\lambda}~ \sqrt{\tfrac{6\pi}{\gamma}} ~ e^{\frac{\omega^2}{8\gamma}}$) as a function of $T$ as $\omega=\arctan \frac{T}{6\lambda}$. Here we have used $\lambda=.1$ and $\gamma=1$. A suitable scaling is used to enlarge the figure.}
\label{norm1}
\end{figure}

Now let us have a closer look at equation (\ref{wnorm}). As $\omega=\arctan \frac{T}{6\lambda}$ so we infer that the norm of the wave packet is time-dependent hence the model is non-unitary. We now investigate the time dependence of the norm ($\tfrac{\pi}{2\lambda}~ \sqrt{\tfrac{6\pi}{\gamma}}~e^{\frac{\omega^2}{8\gamma}}$) as a function of time ($T$) in Figure [\ref{norm1}]. It is observed that with the increase of $T$, the norm flattens, i.e., almost becomes a constant. This indicates that the problem of non-unitarity is somewhat diluted at least for a large time. It deserves mention that this feature may be facilitated by our choice of gauge $n = e^{3\alpha{\beta}_{0}}$ which makes $\dot{T}$ a constant.

\par
\begin{figure}[htb]
\begin{tabular}{c}
\hspace{2.3cm} \includegraphics[width=11cm,height=8cm]{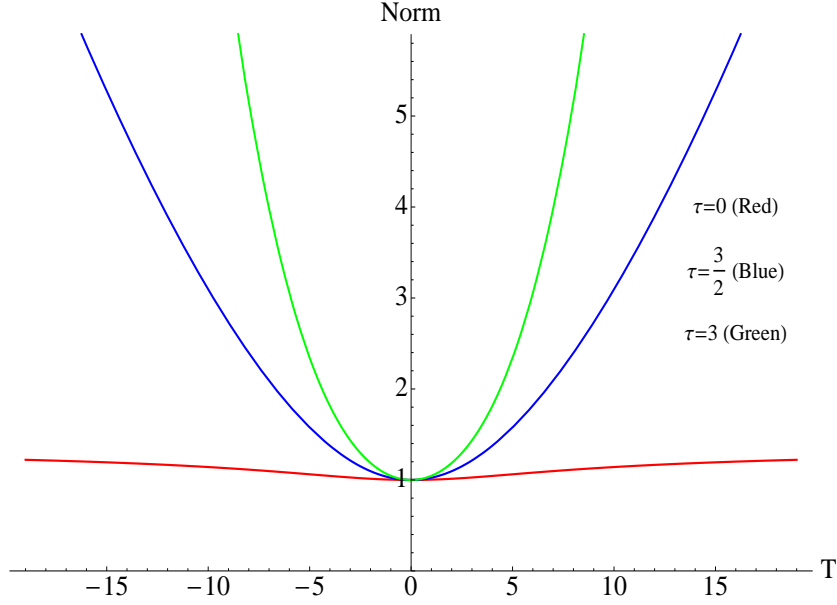}
\end{tabular}
\caption{\scriptsize Plot of Norm $\left(\frac{\pi \sqrt{6\gamma \pi}}{4\gamma} ~
\Gamma \left(1+\frac{\tau}{2} \right) \left(\frac{2}{\lambda}\right)^{1+\frac{\tau}{2}} \left( \lambda^2 + \frac{T^2}{36} \right)^{\frac{\tau}{2}} ~ e^{\frac{\omega^2}{8\gamma}}\right)$ as a function of $T$ as $\omega=\arctan \frac{T}{6\lambda}$. Here we have studied for three different factor orderings $\tau=0~ (Red)$, $\tau=\frac{3}{2}~ (Blue)$ and $\tau=3~ (Green)$. We have used $\lambda=1$ and $\gamma=1$. A suitable scaling is used to enlarge the figure.}
\label{faor}
\end{figure}

It seems important to investigate the same quantum cosmological model with factor orderings other than that mentioned after equation (\ref{wdwe}). The question of factor ordering arises as an artifact of the first term ($-\frac{n}{24} e^{-3\beta_0}~p_0^2$) of the super Hamiltonian because in the quantization process we have operators assigned to each $\beta_0$ and $p_0$. Earlier we have already studied the case where the ordering considered was $e^{-3\beta_0}p_0^2$. We now study two different cases of factor orderings where we consider the factor $e^{-3\beta_0} \times p_0^2$ as
\begin{enumerate}
\item $p_0^2~e^{-3\beta_0}$
\item $p_0~e^{-3\beta_0}~p_0$ .
\end{enumerate}
For the first case we found the solution for the wave function of the Schr$\ddot{o}$dinger-Wheeler-DeWitt equation as
\begin{equation}
\label{comp2}
\Psi_{1^{st}} (\beta_0, \beta_+, T) = e^{3\beta_0}~e^{-iET}~e^{i\tfrac{k}{\sqrt{3}}\beta_+} \big[ C_1 J_{\frac{ik}{2}} \big(\sqrt{36m^2+6E}~e^{2\beta_0}\big)
+ C_2 J_{\frac{-ik}{2}} \big(\sqrt{36m^2+6E}~e^{2\beta_0}~\big)~\big] 
\end{equation}
and for the second case the same entity as
\begin{equation}
\label{comp3}
\Psi_{2^{nd}} (\beta_0, \beta_+, T) = e^{\frac{3}{2}\beta_0}~e^{-iET}~e^{i\tfrac{k}{\sqrt{3}}\beta_+} \big[ C_1 J_{\frac{ik}{2}} \big(\sqrt{36m^2+6E}~e^{2\beta_0}\big)
+ C_2 J_{\frac{-ik}{2}} \big(\sqrt{36m^2+6E}~e^{2\beta_0}~\big)~\big] ~.
\end{equation}
Here $E'^{s}$ and $k'^{s}$ are the separation constants for the respective partial differential equations concerned and $C'^{s}$ are the integration constants. If we compare the equations (\ref{comp1}), (\ref{comp2}) and (\ref{comp3}) we can clearly see that the only difference amongst them is marked by $e^{\tau \beta_0}$ where $\tau =0$ is for (\ref{comp1}), $\tau=3$ for (\ref{comp2}) and $\tau=\frac{3}{2}$ for (\ref{comp3}). So now we consider
\begin{equation}
\Psi (\beta_0, \beta_+, T) = e^{\tau \beta_0}~e^{-iET}~e^{i\tfrac{k}{\sqrt{3}}\beta_+} \big[ C_1 J_{\frac{ik}{2}} \big(\sqrt{36m^2+6E}~e^{2\beta_0}\big)
+ C_2 J_{\frac{-ik}{2}} \big(\sqrt{36m^2+6E}~e^{2\beta_0}~\big)~\big] ~
\end{equation}
as a solution of the Schr$\ddot{o}$dinger-Wheeler-DeWitt equation where $\tau$ is the footprint of different factor orderings considered.
We re-write the super Hamiltonian for the minisuperspace of equation (\ref{supham}) as
\begin{align}
{\cal H} &= {\cal H}_g + {\cal H}_f \nonumber \\ 
&= -\frac{n}{24} e^{-(3+ \varrho + \varsigma)\beta_0}~p_0~e^{\varrho ~\beta_0}~p_0~e^{\varsigma ~\beta_0} \nonumber \\ 
& +n\frac{e^{-3\beta_0}}{24}\big[3\, p_+^2 + 144\, e^{4\beta_0}m^2 + 24\, e^{3(1-\alpha)\beta_0}p_T \,\big] \, .
\end{align}
The relation between $\tau , \varrho$ and $\varsigma$ is given by 
\begin{equation}
\tau = -~ \frac{\varrho~+~2\varsigma}{2} ~,
\end{equation} 
and $(\varrho, \varsigma)$ can take values (0, 0), (0, -3) and (-3, 0). Now we re-examine the earlier results. The wave packet will also be modified and can be expressed as
\begin{equation}
\Psi_{wp} = e^{\tau \beta_0} \sqrt{\pi/\gamma}~ e^\theta ~ e^{i6m^2T} ~ e^{-\frac{e^{4\beta_0}}{4(\lambda + \frac{iT}{6})}} ~ 
e^{-\frac{(\beta_0 + \frac{\beta_+}{\sqrt{3}} + \frac{\theta}{2})^2}{4\gamma}}  ~.
\end{equation}
An explicit calculation for the norm of the wave packet yields
\begin{align}
\label{fonorm}
\int_{-\infty}^\infty  \int_{\infty}^\infty ~ e^{4\beta_0}~ \Psi_{wp}^* ~\Psi_{wp} ~ d\beta_0 ~ d\beta_+ = \frac{\pi \sqrt{6\gamma \pi}}{4\gamma} ~
\Gamma \left(1+\frac{\tau}{2} \right) \left(\frac{2}{\lambda}\right)^{1+\frac{\tau}{2}} \left( \lambda^2 + \frac{T^2}{36} \right)^{\frac{\tau}{2}} ~ e^{\frac{\omega^2}{8\gamma}} ~~.
\end{align}
We can also calculate the expectation values for $\beta_0$ and $\beta_+$ but in that case we get the same results (equation (\ref{e35}) and (\ref{e37})). So the expectation values of the metric functions will not be affected by the factor ordering as we would have the norm of the wave function in the denominator as well ( such as equation (34)). So the physical interpretation remain same. We now look back at equation (\ref{fonorm}). In Figure [\ref{faor}] we show the variation of the norm of the wave packet as a function of time for three different factor orderings.
\par It is well known that the anisotropic models are plagued by the problem of non-unitarity. The problem appears to be generic. But our results show that a clever choice of operator ordering, although unable to solve the problem, can provide a method for alleviating the problem to a certain extent. The operator ordering used in equation (30) ( and hence (35)) does better as figure (2) shows that the norm of the wave packet attains a constant value as $T$ increases. For the two other choices, the expectation value does not lead to a constant value even for high values of $T$ (figure(3)).

\subsection{${\bf \alpha = 1}$ (Stiff Matter)}
With $\alpha= 1$ equation (\ref{vs1}) becomes
\begin{equation}
\frac{\partial^2 \phi}{\partial \beta_0^2} + 144\,e^{4\beta_0}\,m^2 \phi + (24~E  + k^2~) \phi = 0 ~.
\end{equation}
The solution is obtained in terms of Bessel function and we can write
\begin{equation}
\phi = C_3~ J_{\frac{i\sqrt{24E+k^2}}{2}}~ (6m~e^{2\beta_0}~) + C_4~ J_{-\frac{i\sqrt{24E+k^2}}{2}}~ (6m~e^{2\beta_0}~) ~~,
\end{equation}
where $C_3$ and $C_4$ are the integration constants. The wave function can now be written as
\begin{equation}
\Psi (\beta_0, \beta_+, T) = e^{-iET}~e^{i\tfrac{k}{\sqrt{3}}\beta_+} \big[C_3~ J_{\frac{i\sqrt{24E+k^2}}{2}}~ (6m~e^{2\beta_0}~) + C_4~ J_{-\frac{i\sqrt{24E+k^2}}{2}}~ (6m~e^{2\beta_0}~)~\big]~.
\end{equation}

\subsection{${\bf \alpha = \frac{1}{3}}$ (Radiation)}
With $\alpha= 1/3$ equation (\ref{vs1}) becomes 
\begin{equation}
\frac{\partial^2 \phi}{\partial \beta_0^2} + 144\,e^{4\beta_0}\,m^2 \phi + 24 \, E \, e^{2\beta_0} \phi + k^2 \phi = 0 ~.
\end{equation}
The solution can be written in terms of Confluent Hypergeometric function (${\cal F}$) and Associated Laguerre function (${\cal L}$),
\begin{align}
\phi =& e^{-i6me^{2\beta_0}} ~ e^{ik\beta_0} \big[C_5~{\cal F}~ (\tfrac{i(E-im+km)}{2m}, 1+ik, i12me^{2\beta_0}) \nonumber \\
     & + C_6~{\cal L}~ (\tfrac{-i(E-im+km)}{2m}, ik, i12me^{2\beta_0})~\big] ~,
\end{align}
where $C_5$, $C_6$ are the integration constants. The expression for the wave function can be written as
\begin{align}
\Psi (\beta_0, \beta_+, T) =& e^{-iET}~e^{i\tfrac{k}{\sqrt{3}}\beta_+}~e^{-i6me^{2\beta_0}} ~ e^{ik\beta_0} 
\big[C_5~{\cal F}~ (\tfrac{i(E-im+km)}{2m}, 1+ik, i12me^{2\beta_0}) \nonumber \\
 & + C_6~{\cal L}~ (\tfrac{-i(E-im+km)}{2m}, ik, i12me^{2\beta_0})~\big] ~.
\end{align}
\section{Quantization of Bianchi IX cosmological model}
The Bianchi IX metric is written as
\begin{equation}
ds^2 = dt^2 - a^2(t) dr^2 - b^2(t) d\theta^2 - [~ b^2(t) \sin^2 \theta + a^2(t) \cos^2 \theta ~] d\phi^2 + 2 a^2 \cos \theta ~ dr d\phi ~.
\end{equation}
The gravitational Lagrangian density is written as 
\begin{equation}
{\cal L}_g = \frac{2\beta^2 \dot{a}^2}{a^3} - \frac{2\dot{\beta}^2}{a} - \frac{a^5}{2\beta^2} + 2a ~,
\end{equation}
where $\beta=a~b$. The Hamiltonian for this gravitational Lagrangian density becomes
\begin{equation}
{\cal H} = \frac{a^3}{8\beta^2}p_a^2 - \frac{a}{8} p_{\beta}^2 + \frac{a^5}{2\beta^2} - 2a ~,
\end{equation}
where $p_a$ and $p_{\beta}$ are the canonical conjugate momenta for the variables $a$ and $\beta$. Applying Schutz's mechanism \cite{r3,r4} and the
canonical methods described in \cite{r5} we can evaluate the Hamiltonian for the fluid as
\begin{equation}
{\cal H}_f = a^{\alpha} \beta^{-2\alpha} p_T ~.
\end{equation} 
The super Hamiltonian for the minisuperspace of this model is
\begin{align}
{\cal H} &= {\cal H}_g + {\cal H}_f \nonumber \\
&= \frac{a^3}{8\beta^2}~ p_a^2 - \frac{a}{8}~ p_{\beta}^2 + \frac{a^5}{2\beta^2} - 2a + a^{\alpha} \beta^{-2\alpha}~ p_T ~. 
\end{align}
We shall calculate the wave function of the Bianchi IX universe with $\alpha=1$.
\subsection{${\bf \alpha = 1}$ (Stiff Matter)}
Using the quantization procedure as described in \cite{r6,r7} we write the Schr$\ddot{o}$dinger-Wheeler-deWitt equation as
\begin{equation}
-\frac{a^2}{8} \frac{\partial^2 \Psi (a, \beta, T)}{\partial a^2} + \frac{\beta^2}{8} \frac{\partial^2 \Psi (a, \beta, T)}{\partial \beta^2} + 
\frac{a^4}{2} \Psi (a, \beta, T) - 2\beta^2 \Psi (a, \beta, T) + i \frac{\partial \Psi (a, \beta, T)}{\partial T} = 0 ~.
\end{equation}
Now we have to apply the method of separation of variables for the solution of this equation. We write the wave function as
\begin{equation}
\Psi(a, \beta, T) = e^{-iET} \xi (a, \beta)~,
\end{equation}
which leads to
\begin{equation}
-\frac{a^2}{8} \frac{\partial^2 \xi}{\partial a^2} + \frac{\beta^2}{8} \frac{\partial^2 \xi}{\partial \beta^2} + \frac{a^4}{2} \xi - 2\beta^2 \xi
+ E\xi = 0 ~.
\end{equation}
With the separation
\begin{equation}
\xi (a, \beta) = u(a)~v(\beta)~,
\end{equation}
we get
\begin{equation}
\label{pv}
\beta^2 \frac{\partial^2 v}{\partial \beta^2} - 16 \beta^2 v - 8 k v = 0
\end{equation}
and
\begin{equation}
\label{pu}
a^2 \frac{\partial^2 u}{\partial a^2} - 4 a^4 u - 8 (E+k) u = 0 ~.
\end{equation}
Here $k$ is the separation parameter. The solutions of (\ref{pv}) and (\ref{pu}) are known in terms of modified Bessel functions of first (${\cal I}$) and
second (${\cal K}$) kind and can be written as
\begin{equation}
v(\beta) = \beta^{1/2} ~[~ C_7 ~{\cal I}_\nu ~(i4\beta)~ +~ C_8 ~{\cal K}_\nu ~(i4\beta)~]
\end{equation}
and
\begin{equation}
u(a) = a^{1/2} ~[~ C_9 ~{\cal I}_{\nu'}~ (ia^2)~ +~ C_{10} ~{\cal K}_{\nu'} ~(ia^2)~] ~,
\end{equation}
where $\nu = \tfrac{1}{2}\sqrt{1+32k}$ and $\nu' = \tfrac{1}{4} \sqrt{1+32(E+k)}$. $C_i$'s are the arbitrary integration constants. So the final
expression of the wave function is
\begin{align}
\Psi (a, \beta, T) =& e^{-iET}~\sqrt{a\beta}~[~ C_7 ~{\cal I}_\nu ~(i4\beta)~  \nonumber \\
& +~ C_8 ~{\cal K}_\nu ~(i4\beta)~]~[~ C_9 ~{\cal I}_{\nu'}~ (ia^2)~ +~ C_{10} ~{\cal K}_{\nu'} ~(ia^2)~]
\end{align} 
\section{Discussion}
Bianchi type V and type IX perfect fluid models are quantized following Schutz's formalism. As the Wheeler-DeWitt equation for a general equation of state $p=\alpha \rho$ is too involved, some specific examples are taken up. In Bianchi type V models, the solution for the wave function for the universe comes out as a combination of Bessel functions for the case $\alpha = - \frac{1}{3}$. But it deserves mention that the problem of non-unitarity exists in this model, the norm of the wave packet is indeed time dependent. A clever operator ordering can only alleviate the problem in the sense that the norm becomes a constant for a large time. But the problem is only alleviated and not eradicated! In addition to the operator ordering, the choice of a gauge as $n = e^{3\alpha{\beta}_{0}}$ also may have its say. 
\par  The proper volume shows a bounce from a finite minimum and thus the singularity of a zero proper volume can be avoided. By a choice of parameter, the minimum length scale obtained can be tuned to a desired value, say the Planck length. However, this is obtained as a bonus, as the chief motivation of the work had been to check the merits and problems of the particular method of quantization for the anisotropic cosmological models.
\par Only one case, namely $\alpha = - \frac{1}{3}$ has been solved here as an example. This is simply because this case could be integrated anayltically. For some other cases also the wave functions are obtained analytically. But the complete analysis could not be done as the wave packet could not be integrated. So the work is nowhere near a general investigation of the problem. However, so long there is not a complete resolution of the Quantum Cosmology, one has to look for various examples.
\par So although potentially Schutz formalism gives us a beautiful way of handling the problem of the choice of time, it also leads to other problems, namely that of non unitatiry. This could be a generic feature of anisotropic models as indicated in \cite{r11}. The philosophy of interpretation also seems to be awaiting a new direction.

\end{document}